\documentclass[12pt]{article}
\usepackage{graphicx,amsmath,amssymb,euscript,psfrag,epsf}
\usepackage{a4wide}
\newcommand{\be}{\begin{equation}}
\newcommand{\ee}{\end{equation}}
\newcommand{\bea}{\begin{eqnarray}}
\newcommand{\eea}{\end{eqnarray}}
\newcommand{\nl}{\nonumber \\}
\newcommand{\order}{{\cal O}}

\newcommand{\ov}[1]{ \overleftarrow{#1} }

\def\slash#1{#1 \hskip-0.45em /}
\def\Slash#1{#1 \hskip-0.59em /}

\def\beq{\begin{eqnarray}}
\def\eeq{\end{eqnarray}}

\def\eps{\epsilon}

\def\be{\begin{equation}}
\def\ee{\end{equation}}

\def\np{n_+}
\def\nm{n_-}

\def\L{{\cal L}}

\begin{document}

\begin{titlepage}

\begin{flushright}
SI-HEP-2005-02 \\
9.~August 2005
\end{flushright}

\vspace{1.2cm}
\begin{center}
\Large\bf\boldmath
Shape Functions from
 $\bar{B}\to X_c \ell \bar{\nu}_\ell$ \unboldmath 
 \end{center}

\vspace{0.5cm}
\begin{center}
{\sc H. Boos, T. Feldmann, T. Mannel, B.D. Pecjak} \\[0.1cm]
{\sf Theoretische Physik 1, Fachbereich Physik,
Universit\"at Siegen\\ D-57068 Siegen, Germany}
\end{center}

\vspace{0.8cm}
\begin{abstract}
\vspace{0.2cm}\noindent
We study inclusive semi-leptonic $\bar{B}\to X_c \ell \bar{\nu}_\ell$  
decay using the power counting 
$m_c\sim \sqrt{\Lambda_{\rm QCD}m_b}$. 
Assuming this scaling for the charm-quark mass,
the decay kinematics can be chosen to access the shape-function region
even in $b \to c$ transitions.
To apply effective field theory methods in this region we extend 
SCET to describe massive collinear quarks. 
We calculate the tree-level decay rate, including  
${\cal O}(\Lambda_{\rm QCD}/m_b)$ power corrections, 
and show that it factorizes into a convolution
of jet and shape functions.
We identify a certain kinematical variable whose decay spectrum is 
proportional to the universal leading-order shape function familiar
from $b\to u$ decay, 
and speculate as to whether information about this shape function can
be extracted from data on $b\to c$ decay. 

\end{abstract}

\end{titlepage}

\section{Introduction}
The standard tool in heavy quark physics, 
the expansion in inverse powers of the
heavy quark mass, has been extended recently 
to describe the kinematical
situation in which the hadronic jet produced
by the decaying heavy quark carries a large energy of order
$m_b$, but has a small invariant mass of order 
$\sqrt{m_b \Lambda_{\rm QCD}}$. Although processes occurring in this 
so-called shape-function region of phase space do not permit 
a local operator product expansion, they can be described 
by an expansion in terms of  non-local  operators 
evaluated on the light-cone \cite{NeubertSF,BigiSF}. This expansion is most 
conveniently discussed within the framework of
the soft-collinear effective theory (SCET) 
\cite{Bauer:2000ew,Bauer:2000yr,Bauer:2001ct,Beneke:2002ph}, 
as has been done in  studies of soft-collinear factorization in
$b\to u$ transitions  up to one-loop
order \cite{Bauer:2001yt,Bauer:2003pi,Bosch:2004th} and including 
$1/m_b$ power corrections 
\cite{Lee:2004ja,Bosch:2004cb,Beneke:2004in,Tackmann:2005ub}.

Our goal in this work is to examine the effects of final-state 
quark masses on inclusive semi-leptonic $B$ decay occurring in the
shape-function region.  For the up, down, and 
strange quarks, these mass effects are presumably 
very small. This is clearly not the case for the charm quark, 
which is in fact most often treated as a heavy quark with mass  
$m_c\sim m_b$.  In this work 
we suggest  a different power counting for 
the charm-quark mass, namely  
$m_c\sim\sqrt{ \Lambda_{\rm QCD} m_b}$.  This power counting is  
satisfied numerically and allows us to access the shape-function
region even in $b\to c$ transitions \cite{MannelNeubert,MannelTackmann}.
For decay kinematics restricted to the   
shape-function region the charm quark can be
treated as a collinear quark with finite mass. 
It is a simple task to include such a collinear quark
mass in the SCET Lagrangian \cite{Rothstein:2003wh, Leibovich:2003jd}, 
but this has not yet been done beyond leading order 
or applied to a phenomenological example.  

In this paper we formulate a version of SCET 
appropriate for describing processes involving massive
collinear quark fields interacting with soft and collinear gluons,
and use it as a tool to calculate
the hadronic tensor for inclusive semi-leptonic $b\to c$ decay
in the shape-function
region. We calculate the hadronic tensor at tree-level, 
including ${\cal O}(\Lambda_{\rm QCD}/m_b)$ power corrections,
and show that it factorizes into a convolution of 
jet and shape functions. The 
structure of this factorization formula mirrors
closely that of the $b \to u$ case, 
the differences being that a slightly different 
set of shape functions is needed at sub-leading 
order and that the shape functions depend
on a different combination of kinematical variables as compared
with the massless case.  We identify a certain kinematical 
variable whose differential decay 
rate is directly proportional to the universal
leading order shape function,  
and speculate as to whether information on this function
can be extracted from a study of $b\to c$ decay.  
This variable is a generalization of the $P_+$ variable suggested in 
\cite{MannelRecksiegel}, which has been reconsidered in the framework of  
SCET in \cite{Bosch:2004th, NeubertP+}.

The organization of the paper is as follows. 
In the next section we discuss our power counting and 
outline the calculation. We construct the SCET Lagrangian 
for a massive collinear quark in Section \ref{sec:Lagrangian} 
and derive the corresponding transition currents
in Section \ref{sec:Currents}.
In Section \ref{sec:htensor}
we use this version of SCET to calculate the 
hadronic tensor for semi-leptonic $b \to c$
transitions in terms of a factorization formula.
Differential decay distributions are
discussed in Section \ref{sec:spectra}
and compared with the well-known local expansion
of inclusive $b \to c $ spectra in Section \ref{sec:locexpans}. In 
Section \ref{sec:leading}  we discuss possible phenomenological 
applications of our results and we conclude in Section
\ref{sec:conclusions}.

\section{Power Counting and 
Factorization in \bf \boldmath $ \bar B\to X_c \ell \bar \nu_\ell$}

We are interested in calculating decay distributions for 
inclusive $\bar B\to X_c \ell \bar \nu_\ell$ decay in the
shape-function region using effective field theory methods.
In this section we will define our power counting
and explain how a factorization formula arises at tree level.
The central quantity of interest is the hadronic tensor,
which contains all the QCD effects in the semi-leptonic
decay and can be used to derive the differential
decay spectrum.  We define the hadronic tensor as
\be
W^{\mu\nu}=\frac{1}{\pi}{\rm Im}\langle \bar B(v)|
T^{\mu\nu}|\bar B(v)\rangle ,
\ee
where we use the state normalization 
$\langle \bar B(v)|\bar B(v)\rangle=1$. The correlator 
$T^{\mu\nu}$ is given by 
\be\label{eq:correlator}
T^{\mu\nu}=i \int d^4 x e^{-i q\cdot x}{\rm T}
\{J^{\dagger \mu}(x) J^{\nu}(0)\}.
\ee
Here $J^\mu$ is a flavor-changing weak transition 
current, which is
\be
J^\mu=\bar c \gamma^\mu(1-\gamma_5)b
\ee
for $\bar B\to X_c \ell \bar{\nu}_{\ell}$ decay.
We shall use the notation of \cite{DeFazio:1999sv} and
write the hadronic tensor in terms of five scalar structure functions  
\begin{eqnarray} \label{eq:hadtensordef}
W_{\mu \nu} &=& W_1\,  (p_\mu v_\nu + v_\mu p_\nu - g_{\mu \nu} vp 
- i \epsilon_{\mu \nu \alpha \beta} \,p^\alpha v^\beta ) \\ \nonumber
&& - \,W_2 \, g_{\mu \nu} + W_3 \, v_\mu v_\nu 
+ W_4 \, (p_\mu v_\nu + v_\mu p_\nu) + W_5 \, p_\mu p_\nu ,
\end{eqnarray} 
where the independent vectors are chosen to be $v$, the velocity of
the $\bar B$ meson, and $p\equiv m_b v-q$ 
with $m_b$ the $b$ quark pole mass and $q$ the momentum of
the outgoing lepton pair. We use the convention 
$\epsilon^{0123}=-1$.

The hadronic tensor is expressed in the effective theory as a 
double series in the perturbative coupling constant and a
small parameter $\lambda$, which we define through the 
relations
\begin{equation}
\frac{\Lambda_{\rm QCD}}{m_b}  \sim \lambda^2  \quad  \mbox{ and } \quad
\frac{m_c^2}{m_b^2} \sim \lambda^2.
\end{equation}
This choice correlates the two small scales
$m_c/m_b \sim \sqrt{\Lambda_{\rm QCD}/m_b}$.
Quantum fluctuations are described by three widely separated scales,
$m_b^2 \gg m_b^2 \lambda^2 \gg m_b^2 \lambda^4$, provided that
the jet momentum $p$ satisfies $p^2 - m_c^2 \sim  m_b^2 \lambda^2$.  
We will refer to these scales as the hard, jet, and soft scale respectively.
We expect that the fluctuations occurring at these scales 
can be factorized into a convolution of 
hard, jet and soft (shape) functions,  similarly to
the massless case \cite{Bauer:2001yt, Korchemsky:1994jb}.
We will outline the steps used to prove such a factorization formula
to all orders in perturbation theory, under the assumption that there
are no additional complications compared to the massless case, but do
not give a complete proof for decay into charm quarks. 
An essential part of a factorization proof would be to show that
the IR divergences associated with the renormalization of the
leading-power jet function are the same for massive and
massless SCET, and that the IR divergences of the new operators
at sub-leading power can be treated in dimensional regularization.
Our aim here is to explain how our explicit tree-level 
results can be interpreted as the leading-order term of such a 
presumed all-orders factorization formula.

We calculate the hadronic tensor by means 
of a two-step matching procedure,
${\rm QCD}\to {\rm SCET}\to {\rm HQET}$.
In the first step, we remove fluctuations at the hard scale
by matching the QCD Lagrangian and transition current onto 
their corresponding expressions in SCET.
We derive the SCET Lagrangian in Section \ref{sec:Lagrangian}
and discuss the transition currents in 
Section \ref{sec:Currents}, to relative order $\lambda^2$ in
the power counting.  In addition to these SCET
expressions, we also need the sub-leading HQET
Lagrangian 
\be
\L^{(2)}_{\rm HQET}=\frac{1}{2m_b}\left[\bar h_v (iD_s)^2 h_v 
+ g \frac{C_{\rm mag}}{2}\bar h_v\sigma_{\mu\nu}F_s^{\mu\nu}h_v\right].
\ee 
In general, $C_{\rm mag}\ne 1$ due to fluctuations at the hard
scale, and the  SCET transition currents 
are multiplied by Wilson coefficients depending on quantities
at the hard scale.
These Wilson coefficients define the
hard functions in the factorization formula. However,  
the SCET transition currents and HQET Lagrangian have trivial
hard coefficients at tree level, so we will not need to discuss
these hard functions any further.  

After this first step of matching, the correlator 
(\ref{eq:correlator}) is  calculated using the effective-theory 
expressions for the Lagrangian and currents.  
This will be the subject of Section \ref{sec:htensor}. 
The effective theory contains a collinear charm-quark 
field and an HQET field for the 
$b$ quark, as well as soft
and collinear light quarks and gluons.  
In the next section we will define
more precisely what we mean by collinear and soft, for now
we simply note that the collinear fields fluctuate 
at the scale $m_b^2 \lambda^2$, whereas the soft and HQET fields
fluctuate at the scale $m_b^2 \lambda^4.$   
Since the $B$ meson contains only soft degrees of freedom, 
the matrix element of each term in 
the correlator factorizes into the form
\bea\label{eq:factorization}
\langle\bar B| T^{\rm eff}_i|\bar B\rangle&=& \langle\bar{B}
|\bar{h}_v[{\rm soft\, fields}]h_v
|\bar{B}\rangle \otimes
\langle 0 |[{\rm collinear\, fields}] |0\rangle \nonumber \\
&\equiv & S_i\otimes J_i  \,  ,
\eea 
where the $\otimes$ stands for a convolution.
The vacuum matrix
element of the collinear fields defines a set of perturbatively
calculable jet functions $J_i$ containing fluctuations at the
jet scale. The tree-level jet functions are given in terms
of propagators of the collinear fields.  Calculating
these jet functions removes the
collinear degrees of freedom and defines the second
step of matching ${\rm SCET} \to{\rm HQET}$.
The matrix element of the soft fields between the 
$\bar B$ meson states is calculated in HQET 
and defines a set of non-perturbative
shape functions $S_i$.  The differential decay distributions
are then written in terms of the $W_i$, each of which
factorizes into a convolution of jet and shape functions.

\section{Matching onto the SCET Lagrangian}\label{sec:Lagrangian}

In this section we derive the
SCET Lagrangian for soft and collinear gluons interacting
with collinear charm-quark fields carrying a mass $m_c \sim \lambda m_b$.  
The leading-order result
has been given previously in \cite{Leibovich:2003jd}.
Here we extend the derivation to sub-leading order,
using the position space formalism of \cite{Beneke:2002ph}.  
We do not discuss the Yang-Mills Lagrangian, because
its matching onto the effective theory is the 
same as in \cite{Beneke:2002ni}.

The  starting point is the QCD Lagrangian for massive quark fields,
\be\label{eq:QCDL}
\L_{\rm QCD}=\bar \psi (i \Slash{D}-m_c) \psi,
\ee
where $iD^\mu=i\partial^\mu + g A^\mu$.   
The derivation of the effective theory proceeds in the 
usual way.    
We introduce effective theory fields according 
to the momentum scaling of their Fourier components.  
In terms of two light-like vectors
$\np,\nm$ satisfying $\np \nm =2$, the components
of a collinear momentum $p_c^\mu$ scale as $(\np p_c,p_{c\perp},\nm p_c)\sim
m_b(1,\lambda,\lambda^2)$ and those of a soft momentum
as $p_s^\mu$ as $\sim m_b (\lambda^2,\lambda^2,\lambda^2)$.

Following the standard procedure, we
decompose  the QCD gluon field into a collinear field $A_c^\mu$ and a 
soft contribution $A_s^\mu$. With our power counting for the
jet momentum, we only have to consider charm-quark momenta
which satisfy $p_c^2-m_c^2\sim m_b^2 \lambda^2$.
These are described by massive collinear quark fields. 
Because of the finite charm-quark mass, 
loop diagrams do not contain soft divergences related to
charm propagators. Therefore charm quarks appear neither
in the soft Lagrangian nor in the $B$\/-meson states.
This is in contrast to the $b \to u$ decay, where the
effective Lagrangian contains interactions between
soft and collinear light-quark fields.

Notice that in the region of phase space where
$p_c^2-m_c^2\sim m_c \Lambda_{\rm QCD} \sim  m_b^2 \lambda^3$,
we would have to consider another effective theory which describes
exclusive hadronic charm resonances.
Since $\Lambda/m_c\sim \lambda$
is a small quantity, one can switch to HQET and
split the charm-quark
momentum into $p_c = m_c v^\prime + k^\prime$, where $k^\prime$
is a residual momentum. Because the energy transfer
to the charm quark is still large, it is boosted by a
factor of ${\cal O}(1/\lambda)$ with respect to the
$B$\/-meson rest frame. Therefore, the charm velocity
scales like $v^\prime\sim (1/\lambda, 1,\lambda)$,
and the residual momentum scales as $k^\prime\sim
m_b(\lambda,\lambda^2,\lambda^3)$.
One can then
introduce an effective theory field $h_{v^\prime}$ 
carrying a dynamical momentum $k^\prime$, which is described 
in terms of a ``boosted HQET''. 
Restricting ourselves to collinear quark fields, we have
to make sure that the considered phenomenological observables
are not sensitive to the exclusive region, see the discussion
in Section~\ref{sec:leading}.

For the collinear 
quark fields we define the projections 
\be
\xi(x)= \frac{\Slash{n}_- \Slash{n}_+}{4}\psi_c(x), \qquad \eta(x) = 
\frac{\Slash{n}_+ \Slash{n}_-}{4}\psi_c(x).
\ee
Inserting these into (\ref{eq:QCDL}) and solving the field
equation for $\eta$, one finds 
\be \label{eq:etafield}
\eta = -\frac{\slash{n}_+}{2}\frac{1}{i\np D}
(i\Slash{D}_\perp -m_c)\xi,
\ee
so that the Lagrangian becomes
\bea\label{eq:lagrangian}
{\cal L}_{\rm QCD} = 
\bar \xi \left[i\nm D +(i\Slash{D}_\perp -m_c)\frac{1}{i\np D}
(i\Slash{D}_\perp + m_c)\right]
\frac{\slash{n}_+}{2}\xi.
\eea
The manipulations performed so far are just a rewriting of the
QCD Lagrangian. In the next step we expand this
Lagrangian as a series in the parameter $\lambda$.  
The expansion is related to the soft gluon fields
contained in the covariant derivative 
$i D^\mu = i\partial^\mu + g A_c^\mu +  g A_s^\mu$,
as well as the multi-pole expansion of these soft fields. 
The result for the leading-order  Lagrangian is
\bea\label{eq:leadingL}
{\cal L}^{(0)}_{\xi}&=& \bar\xi
\left(i\nm D + (i \Slash{D}_{\perp c}-m_c)\frac{1}{i\np D_c}
(i \Slash{D}_{\perp c}+m_c)\right)\frac{\slash{n}_+}{2}\xi,
\eea
while the sub-set of power-suppressed terms proportional to the
mass is
\bea\label{eq:subLm}
\L_{\xi m }^{(1)}&=&m_c\bar\xi
[g \Slash{A}_{\perp s},\frac{1}{i\np D_c}]\frac{\slash{n}_+}{2}
\xi,\nonumber \\
\L_{\xi m 1}^{(2)}&=& m_c^2 \bar\xi \frac{1}{i\np D_c} g\np A_s
\frac{1}{i \np D_c}\frac{\slash{n}_+}{2}\xi, \nonumber \\
\L_{\xi m 2}^{(2)}&=& m_c \bar\xi[\frac{1}{i\np D_c} g \np A_s
\frac{1}{i\np D_c}, i \Slash{D}_{\perp c}]\frac{\slash{n}_+}{2}\xi,
\nonumber \\
\L_{\xi m 3}^{(2)}&=&m_c\bar\xi
[(x_\perp\partial_\perp g \Slash{A}_{\perp s}),
\frac{1}{i\np D_c}]\frac{\slash{n}_+}{2}
\xi,
\eea
where ${\cal L}^{(1)}$ and ${\cal L}^{(2)}$ contribute to relative
order $\lambda$ and $\lambda^2$ to the effective action.
In the terms above the collinear fields are evaluated at $x$, but
the soft fields are multi-pole expanded and depend
only on $x_-^\mu=(\np x/2)\nm^\mu \equiv x_+ \nm^\mu$.
We will not make this explicit in the following, where
it is always understood that
\bea
&&  h_v(x) \equiv h_v(x_+ n_-^\mu) \ , \qquad
    A_s(x) \equiv A_s(x_+ n_-^\mu) \ .
\label{multi}
\eeq
The remaining power-suppressed terms are the same as in the massless
case. One can still make the standard field redefinitions
($\xi= Y \xi^{(0)}$, etc.) to decouple the soft
gluon field from the leading-order collinear Lagrangian
(\ref{eq:leadingL}), a fact
which is crucial to factorization.  
 
The Lagrangian can be written in a manifestly gauge-invariant
form by redefining the collinear fields
using the methods developed in \cite{Beneke:2002ni}.
Applying these techniques, we find that the result
for the leading-order Lagrangian (\ref{eq:leadingL}) 
is unchanged, but
to relative order $\lambda^2$ the power suppressed terms  
are the same as in the massless case and read \cite{Beneke:2002ni} 
\bea
\label{eq:lagrangians}
{\cal L}_\xi^{(1)}&=&\bar{\xi}x_{\perp}^\mu n_-^\nu W_c \,
g F^s_{\mu\nu} W_c^\dagger
\frac{\slash{n}_+}{2}\xi \nonumber \\
{\cal L}^{(2)}_{1\xi } & = & \frac12\bar{\xi} (\nm x)  n_+^\mu n_-^\nu 
W_c \,g F^s_{\mu\nu} W_c^\dagger\frac{\slash{n}_+}{2}
\xi, 
\nonumber 
\\
{\cal L}^{(2)}_{2\xi } &=& \frac12\bar{\xi}x_\perp^\mu x_{\perp\rho}n_-^\nu
W_c[D_{\perp s}^{\rho},g
F_{\mu\nu}^s]W_c^\dagger\frac{\slash{n}_+}{2}\xi, 
\nonumber 
\eea
\bea
{\cal L}^{(2)}_{3\xi } &=&\frac{1}{2} \bar{\xi}  i\Slash{D}_{\perp c}
\frac{1}{i\np D_c}
x_{\perp}^\mu \gamma_{\perp}^\nu W_c \,g F^s_{\mu\nu}W_c^\dagger
\frac{\slash{n}_+}{2}\xi \nonumber \\
&&+\,\frac12\bar{\xi}x_\perp^\mu\gamma_{\perp}^{\nu}
W_c \,g F^s_{\mu\nu}W_c^\dagger\frac{1}{i\np D_c}
i\Slash{D}_{\perp c}\frac{\slash{n}_+}{2}\xi.
\eea
The power-suppressed mass terms first appear at  
relative order  $\lambda^3$. To see this, we note that after making
the field redefinitions and working in the collinear
light-cone gauge $n_+ A_c =0$, one effectively replaces
\be
\frac{1}{i\np D}\to \frac{1}{i \np \partial}+{\cal O}(\lambda^3)
\ee 
in the expansion of (\ref{eq:lagrangian})
(see  eq.~(23) of \cite{Beneke:2002ni}). It is then easy 
to show that all mass terms not contained in the leading-order 
Lagrangian are suppressed by at least a factor
$\lambda^3$. 

While no explicit power-suppressed mass effects
appear to $\order (\lambda^2)$ accuracy in the
effective Lagrangian
after making the field
redefinitions, we will see later on that they re-appear
in the time-ordered products of SCET Lagrangians and currents.
The calculation with either form of the Lagrangian and currents
yields the same result, 
namely the appearance of one new sub-leading shape function
as compared to the massless case.

\section{Effective theory  currents}\label{sec:Currents}
In this section we consider the matching of the flavor-changing
weak transition current defined in QCD onto its corresponding
expression in SCET.  For $b\to c$ transitions, this matching
takes the form
\be
(\bar\psi_c \Gamma b)_{\rm QCD}\to e^{-im_b v x}
\left[J^{(0)} + J^{(1)}+ J^{(2)}+...\right],
\ee
where $\bar \psi_c$ is a collinear charm-quark field.
In general, the $J_i$ are convolutions of a short distance
Wilson coefficient with operators built out of SCET and
HQET fields. In what follows, however, we will work only at 
tree level, where the convolutions reduce
to simple multiplication. The power-suppressed
currents we need here can be deduced from the
results in \cite{Beneke:2002ni} by
noting that at tree level all powers of the mass 
are related to  the $\eta$ field. 
Using  (\ref{eq:etafield}) in conjunction with
\cite{Beneke:2002ni}, we find for the mass terms   
\bea
J^{(1)}_m &=& m_c\, \bar\xi  \frac{\slash{n}_+}{2}
\frac{1}{-i\np\ov D_c}W_c\Gamma h_v,\label{eq:J1m}\\
J^{(2)}_{m1} &=& m_c \, \bar\xi \frac{\slash{n}_+}{2}
\frac{1}{-i\np\ov D_c}W_c \Gamma x_\perp D_{\perp s} h_v, \\
J^{(2)}_{m2} &=& m_c \, \bar\xi \frac{\slash{n}_+}{2}
\frac{1}{i\np\ov D_c}\Gamma\frac{\slash{n}_-}{2m_b}
[i \Slash{D}_{\perp c} W_c]  h_v.
\eea
The convention is such that derivatives do not act outside of 
the square brackets, and the soft gluon field
and the heavy quark field are multi-pole
expanded. We have again used the field redefinitions of 
\cite{Beneke:2002ni} to
write the result in a manifestly
gauge invariant form. The remaining terms are the same as 
in the massless case \cite{Beneke:2002ni, Chay:2002vy, Pirjol:2002km}.  
The sub-set of these that will be needed later
on when calculating the tree-level time-ordered products is
\begin{eqnarray}
\label{eq:currents}
J^{(0)}&=&\bar{\xi} W_c \Gamma  h_v,   \nonumber\\
J^{(1)}_{1}& =& \bar{\xi} W_c \Gamma  x_{\perp\mu}D_{\perp s}^{\mu}h_v, 
\quad 
J^{(1)}_{2}=-\bar{\xi} \,i\Slash{{\ov D}}_{\!\perp c }\frac{1}{i
  \np\ov D_c} W_c \frac{\slash{n}_+}{2}\Gamma  h_v,\nonumber\\
J^{(2)}_{1}&=&\bar{\xi} W_c \Gamma \frac{\nm x}{2}\,\np D_s h_v, \quad 
J^{(2)}_{2}=\bar{\xi} W_c \Gamma \frac{x_{\mu\perp}x_{\nu\perp}}{2}
D^\mu_{\perp s}\,D^\nu_{\perp s}h_v,\nonumber\\
J^{(2)}_{3}&=&-\bar{\xi}\,i\Slash{{\ov D}}_{\!\perp c}
\frac{1}{i {\np\ov D_c}} W_c
\frac{\slash{n}_+}{2}\Gamma x_{\perp\mu}D^\mu_{\perp s}h_v, \quad 
J^{(2)}_{4}=\bar{\xi} W_c \Gamma \,\frac{i{\Slash{D}}_s}{2m_b}h_v.
\end{eqnarray}

\section{The hadronic tensor at tree level}
\label{sec:htensor}

In this section we obtain an expression for the hadronic tensor
in terms of a factorization formula. We work at tree level
and include ${\cal O}(\lambda^2)$ power corrections. We start by
calculating the time-ordered products appearing in  the 
correlator (\ref{eq:correlator}) using the effective theory
Lagrangian and currents derived in the previous sections,
setting $A_c  = 0$ as appropriate at tree level. 

Many time-ordered products are possible at ${\cal O}(\lambda^2)$.
It is convenient to divide these into 
two groups: those that vanish when $m_c\to 0$, and those that 
do not. Since the sub-leading Lagrangian (\ref{eq:lagrangians}) in its 
gauge-invariant form is actually the same as in the 
massless case, we can identify the time-ordered 
products which do not vanish when
$m_c\to 0$ from eq.~(54) of \cite{Beneke:2004in}.  
For these terms the difference between the two cases lies 
only in the 
leading-order Lagrangian, and we can account for this
difference  by replacing the propagator for
massless collinear quark fields by that for massive fields,
which is
\bea
\langle 0|{\rm T}\{\xi(x)_{a\alpha}
\bar \xi(y)_{b\beta}\}|0\rangle & \equiv & i\Delta(x-y)\delta_{ab}
\left(\frac{\slash{n}_-}{2}\right)_{\alpha\beta} \nonumber\\
&=&\int \frac{d^4 k}{(2 \pi)^4}\, e^{-i k(x-y)}
\frac{i\np k}{k^2 -m_c^2+ i\eps}
\left(\frac{\slash{n}_-}{2}\right)_{\alpha\beta}
\delta_{ab}.
\eea
The Latin (Greek) indices refer to color (Dirac) indices.
The function $\Delta(z)$ satisfies
\be\label{eq:eqmotion}
i\nm\partial\Delta(z)=\delta^{(4)}(z)+\frac{1}{i\np \partial}
\left(m_c^2-(i\partial_\perp)^2\right)\Delta(z).
\ee
We will always work in the frame where $p_\perp =0$, such
that in tree-level expressions $\partial_\perp$ can be dropped. 

\begin{figure}[t]
\begin{tabular}{c}
\hspace{0cm}\includegraphics[width=1\textwidth]{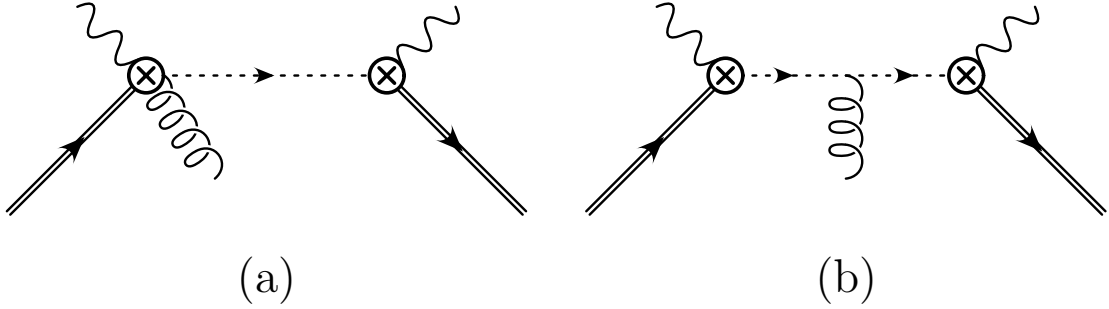} 
\end{tabular}
\caption{
Representative Feynman diagrams involving soft gluon emission. These
give rise to (a) bi-local and (b) tri-local  shape functions.
\label{figure}}
\end{figure}

At tree level
the factorization formula contains up to a double convolution between
the jet and shape functions. The tree-level jet functions
are related to (products of)
propagators, and some of the convolution integrals can be 
performed explicitly, after which the results can be 
written as a convolution over a single variable only. To show
how this works and to shorten intermediate expressions, 
we introduce the integral operators 
($m_c^2\equiv m_c^2-i\eps, \np p>0$)
\bea \label{eq:I2}
I_2 * f  &\equiv& -\int d^4 x e^{ipx}\int \frac{d^4 k}{(2\pi)^4}
e^{-ikx}\frac{\np k}{k^2-m_c^2 }f(x_+)\nl
&=&  -\int dx_+ e^{i\nm p x_+}\int \frac{d\nm k}{2\pi}
e^{-i\nm k x_+}\frac{1}{\nm k -m_c^2/\np p}f(x_+)\nonumber\\
&=&i \int_0^\infty dx_+ e^{i(\nm p -m_c^2/\np p)x_+} f(x_+), 
\eea
\bea \label{eq:I3}
I_3 * g &\equiv&
-\int d^4 x d^4 z e^{ipx}\int \frac{d^4 k}{(2\pi)^4}
\frac{d^4 k^\prime}{(2\pi)^4} e^{-ik(x-z)}e^{-ik^\prime z}
\frac{\np k}{(k^2 -m_c^2)
(k^{\prime 2}-m_c^2)}g(x_+,z_+)\nl
&=&-\int dx_+ dz_+ e^{i\nm p x_+}\int \frac{d\nm k}{2 \pi}
\frac{d\nm k^\prime}{2\pi}e^{-i\nm k(x_+-z_+)}e^{-i\nm k^\prime z_+ }
\nonumber \\ &&
\hspace{2cm}\frac{1}{\np p}\left(\frac{1}{\nm k -m_c^2/\np p}\right)
\left(\frac{1}{\nm k^\prime - m_c^2/\np p}
\right)g(x_+,z_+) \nonumber \\
&=&I_2 * \frac{-i}{\np p}\int_0^{x_+} dz_+ \, g(x_+,z_+).
\eea
In terms of the factorization formula, the functions 
$f$ and $g$ represent generic shape functions depending
on one or two variables respectively,
and the momentum-dependent propagators correspond to
tree-level jet functions. These are represented by  Feynman
diagrams such as Figure \ref{figure}(a) for $f(x_+)$, 
and Figure \ref{figure}(b) for $g(x_+,z_+)$. 
Performing the integral to arrive at the final lines
of (\ref{eq:I2}, \ref{eq:I3}) is equivalent
to performing a convolution integral
in the factorization formula.    In (\ref{eq:I2}) this is trivial,
but in (\ref{eq:I3}) the object 
$\int_0^{x_+} dz_+ \, g(x_+,z_+)$ introduces
an ``effective shape function'' of a single variable only.  
To keep the structure of factorization as presented in
(\ref{eq:factorization}) clear, we will be careful  
to distinguish these effective shape functions from
the shape functions defined directly by the multi-local
matrix elements of soft operators between $\bar B$ meson states.

The integral operators $I_2,\,I_3$
are the same as in \cite{Beneke:2004in}
after replacing $\nm p\to \nm p-m_c^2/\np p$ in the 
exponential of $I_2^*$ defined there.  
We can thus obtain the part of the correlator that does not
vanish when $m_c\to 0$ from 
eq.~(54) of that work by making this replacement, and leaving
out the term involving the soft quark field, 
which does not appear in the 
massive case.  We also 
need to identify a small number of terms which
vanish when $m_c\to 0$ and  cannot 
be derived from previous results. 
First, we need insertions of the transition 
currents $J_m$.  Using the short-hand notation
\be
J_A^\dagger\, J_B\, i {\cal L}_C \equiv i\int d^4 x e^{ipx}\,T\left\{
J_A^\dagger (x)J_B(0) i\int d^4 z {\cal L}_C(z)\right\},
\ee
and working in soft light-cone gauge $n_- A_s =0$,
the relevant time-ordered products are 
 \bea
J^{(0)^\dagger}J^{(1)}_m+J^{(1)^\dagger}_m J^{(0)}&=&
\frac{m_c}{\np p} I_2 *
\bar h_v(x)  \bar\Gamma \Gamma h_v(0),\\
J^{(1)\dagger}_m J^{(1)}_m &=&\frac{m_c^2}{(\np p)^2}I_2 *
\bar h_v(x)  \bar\Gamma \frac{\slash{n}_+}{2}\Gamma h_v(0),\\
J^{(2)\dagger}_{m1} J^{(0)} = 0.
\eea
(Recall that the soft fields are multi-pole expanded, see (\ref{multi}).)
The only additional mass terms are related to insertions
of
\be
\L_{1\xi}^{(2)}= \frac12 \bar\xi\, (\nm x) \, \np^\mu \nm^\nu W_c
g F_{\mu\nu}^s W_c^\dagger \frac{\slash{n}_+}{2}\xi,\quad
J_1^{(2)}=\bar\xi\Gamma \frac{\nm x}{2}\np D_s h_v.
\ee
They contribute to the correlator through the time-ordered products
\bea\label{eq:sum1}
 J_1^{(2)\dagger}J^{(0)}+J^{(0)\dagger}J^{(0)}i\L_{1\xi}^{(2)}
& = &
-\frac{m_c^2}{(\np p)^2}
I_2 * (i x_+)
\bar h_v(x) \, i\np\ov\partial\bar\Gamma
\frac{\slash{n}_-}{2}\Gamma h_v(0)\nonumber \\
&& -\frac{m_c^2}{\np p}
I_3 *\bar h_v(x) \, g\np A_s(z) \bar\Gamma
\frac{\slash{n}_-}{2}\Gamma h_v(0) \nl
&=&
\frac{i m_c^2}{(\np p)^2}I_2 * \int_0^{x_+}dz_+
\bar h_v(x)(-i\np {\ov D}_s)(z) \bar\Gamma
\frac{\slash{n}_-}{2}\Gamma h_v(0).
\cr &&
\eea
We used an integration by parts to write the field
strength tensor in terms of the soft gluon field and then
applied (\ref{eq:eqmotion}) to simplify (\ref{eq:sum1}).
The simple form of the tree-level propagators has allowed us to
replace $\nm x$ with $\np x \,  m_c^2/(\np p)^2$ in this equation, as is
easily verified by writing the factors of $x$ in momentum space
as derivatives acting on tree-level propagators.
These expressions can be made 
gauge invariant by inserting the appropriate 
Wilson lines.  After including these Wilson lines, the tree-level
result for the correlator can be written entirely in terms of
multi-local covariant derivatives of the type in (\ref{eq:sum1}).  
In an arbitrary gauge these are defined by
\be
(Y^\dagger i D_s^\mu Y)(z)_{ab}=
i\partial^\mu\delta_{ab} + (Y^\dagger [i D_s^\mu Y])(z)_{ab}.
\label{YDs}
\ee
 The indices refer to color and the derivative does not
act outside the square brackets. At tree level the result 
for the correlator 
with arbitrary Dirac structure is
\begin{eqnarray}
&& T^{\mu\nu}(u) = -\int\frac{d x_+ d\nm k}{2\pi} \,
e^{i (u -\nm k) x_+}\,\frac{1}{\nm k+i\epsilon}\,\times
\bigg\{ \nonumber \\
&& \hspace*{1cm}(\bar h_v Y)(x)\bar\Gamma 
\left[\frac{\slash{n}_-}{2}+\frac{m_c}{\np p}
+\frac{m_c^2}{(\np p^2)}\frac{\slash{n}_+}{2}\right] \Gamma 
(Y^\dagger h_v)(0)
\nonumber\\
&& \hspace*{1cm}
+T\Big\{(\bar h_v Y)(x)\bar\Gamma \frac{\slash{n}_-}{2}\Gamma 
(Y^\dagger h_v)(0)\,
\,i\int d^4 z\,{\cal L}_{\rm HQET}^{(2)}(z)\Big\}
\nonumber\\
&&  \hspace*{1cm}
+\,\frac{1}{2 m_b} \Big[
(\bar h_v (-i\overleftarrow{\Slash{D}}_s)Y)(x)\bar\Gamma 
\frac{\slash{n}_-}{2}\Gamma (Y^\dagger h_v)(0) + 
(\bar h_v Y)(x)\bar\Gamma \frac{\slash{n}_-}{2}\Gamma 
(Y^\dagger i\Slash{D}_s h_v)(0)\Big]
\nonumber\\
&&  \hspace*{1cm} -\, \frac{1}{n_+ p}\,
\Big[(\bar h_v  (-i\overleftarrow{D}_s^{\mu_\perp})Y)(x)\bar\Gamma 
\frac{\slash{n}_+}{2}\gamma_{\mu_\perp}\!
\frac{\slash{n}_-}{2}\Gamma (Y^\dagger h_v)(0) \nonumber\\[-0.2cm]
&&\hspace*{4cm} +\, 
(\bar h_v Y)(x)\bar\Gamma
\frac{\slash{n}_-}{2} \gamma_{\mu_\perp}\!\frac{\slash{n}_+}{2}\Gamma 
(Y^\dagger (-i \overleftarrow{D}_s^{\mu_\perp})h_v)(0)\Big]
\nonumber\\
&&  \hspace*{1cm} +\, \frac{i}{n_+ p}\,
\int^{x_+}_0 \!d z_+(\bar h_v Y)(x)\bar\Gamma \frac{\slash{n}_-}{2}
(Y^\dagger (-i \overleftarrow{\Slash{D}}_{s\perp}) 
(-i \overleftarrow{\Slash{D}}_{s\perp})  Y)(z)\Gamma (Y^\dagger h_v)(0)
\nonumber\\
&&  \hspace*{1cm} +\, \frac{im_c^2}{(\np p)^2} 
\int^{x_+}_0 \!d z_+(\bar h_v Y)(x)
(Y^\dagger (-i \np \overleftarrow{D}_{s})Y)(z)
\bar\Gamma \frac{\slash{n}_-}{2}\Gamma (Y^\dagger h_v)(0)
\bigg\},
\label{eq:final1}
\end{eqnarray}
where we have defined $u\equiv \nm p-m_c^2/\np p$.

In order to calculate the hadronic tensor, we need to take the imaginary
part of the correlator and  decompose the matrix elements between
$\bar B$ meson states in terms of scalar functions.  
The matrix elements of the operators on the final two lines 
of (\ref{eq:final1}) could be used to directly define a set
of effective shape functions depending on the variable
$u$. To keep factorization explicit we do not take this approach,
and introduce effective shape functions only in terms
of convolutions involving
multi-local shape functions defined through $\bar B$
meson matrix elements. After contracting with the color
indices of the tree-level jet functions, we need the following
matrix elements (the Greek subscripts indicate Dirac indices
and $\eps_{\mu\nu}^\perp\equiv i
\eps_{\mu\nu\rho\sigma}n_-^\rho v^\sigma$)
\bea\label{eq:shapes}
&&\langle
\bar B |(\bar h_v Y)(x)_{\alpha}(Y^\dagger h_v)(0)_\beta |\bar B \rangle
= \frac12 \left(\frac{1+\slash{v}}{2}\right)_{\beta\alpha}
{\tilde S}(x_+),
\\
&&\langle\bar B |(\bar h_v Y)(x)_{\alpha}(Y^\dagger h_v)(0)_\beta 
\,i\int d^4 z \, \L^{(2)}_{\rm HQET }(z) |\bar B\rangle=
\frac{1}{2m_b}\frac12 
\left(\frac{1+\slash{v}}{2}\right)_{\beta\alpha}
{\tilde s}(x_+),\nonumber \eea
\bea
\label{35}
&&\langle \bar B|(\bar h_v Y)(x)_{\alpha}(Y^\dagger
i D_s^{\mu} Y)(z)(Y^\dagger h_v)(0)_{\beta}|\bar B\rangle = \\
&&\frac12 \left(\frac{1+\slash{v}}{2}\right)_{\beta\alpha}
\left[-i{\tilde S}^{\prime}(x_+)v^\mu + 
\left(i{\tilde S}^{\prime}(x_+)-
{\tilde T}_1(x_+,0)+{\tilde T}_1(x_+,z_+)\right)n_-^\mu\right]\nonumber\\
&&+\frac{\eps_\perp^{\mu\rho}}{4}\left(\frac{1+\slash{v}}{2}
\gamma_{\rho_\perp}
\gamma_5\frac{1+\slash{v}}{2}\right)_{\beta\alpha}
\left[{\tilde t}(x_+)-{\tilde T}_2(x_+,0)
+{\tilde T}_2(x_+,z_+)\right],\nonumber \\
&&\langle \bar B|(\bar h_v Y)(x)_{\alpha}(Y^\dagger i D^s_{\mu_\perp}
i D^s_{\nu_\perp} Y)(z)(Y^\dagger h_v)(0)_{\beta}|\bar B\rangle = \\
&& \frac{g_{\mu\nu}^\perp}{4}
\left(\frac{1+\slash{v}}{2}\right)_{\beta\alpha}
\bigg[{\tilde u}_1 (x_+)+{\tilde U}_1(x_+,z_+)\bigg]-
\frac{\eps^\perp_{\mu\nu}}{4}
\left(\frac{1+\slash{v}}{2}\slash{n}_-
\gamma_5\frac{1+\slash{v}}{2}\right)_{\beta\alpha}
{\tilde U}_3(x_+,z_+).\nonumber
\eea
The results are written in momentum space using
the convention
\be
\tilde S_i(x_{1+},\dots, x_{n +})=\int d\omega_1\dots d \omega_n
e^{-i(\omega_1 x_{1+}\dots +\omega_n x_{n+})}S_i(\omega_1,\dots,\omega_n)
\ee
for the Fourier transform of a generic
shape function $S_i$. We now introduce effective shape functions
defined by the convolutions
\begin{eqnarray}\label{eq:effshape}
u_s(u)&=& \int d\omega_1 d\omega_2 \,J_2(u;\omega_1,\omega_2)
\,[u_1(\omega_1)\delta(\omega_2)+U_1(\omega_1,\omega_2)],
\nonumber\\
u_a(u)&=& \int d\omega_1 d\omega_2 \,J_2(u;\omega_1,\omega_2) 
\,U_3(\omega_1,\omega_2), \nl
t_1(u)&=& \int d\omega_1 \, J_2(u;\omega_1,0) \left[\omega_1 S(\omega_1)
-2 \int d\omega^\prime T_1(\omega_1,\omega^\prime)\right]
\nl
&& + 2\int d\omega_1 d\omega_2\, J_2(u;\omega_1, \omega_2)\,
 T_1(\omega_1,\omega_2);
\end{eqnarray}
\begin{eqnarray}
J_2(u;\omega_1,\omega_2)&=&
-\frac{1}{\pi}\,{\rm Im}\frac{(\np p)^2}{(p_{\omega_1}^2-m_c^2)
( p_{\omega_{12}}^2-m_c^2)} \nl
&=&\frac{1}{\omega_2}\big(
\delta(u -\omega_1-\omega_2)-\delta(u-\omega_1)\big),
\end{eqnarray}
where 
$p_{\omega 12\dots j}=p-\omega_{12 \dots j}\np /2$ 
and $\omega_{12\dots j}=\omega_1+\omega_2 +\dots \omega_j$.
$J_2$ is the tree-level jet function related to  the product
of propagators in (\ref{eq:I3}).

The contributions to the hadronic tensor
which do not vanish when $m_c\to 0$ can be obtained from
eq.~(59) of \cite{Beneke:2004in} by replacing 
$\nm p \to u$. 
The terms which vanish when $m_c\to 0$ are
\bea
 W^{\mu\nu}_m &=&
\left(\frac{m_c}{ \np p}\frac12 {\rm tr}
\left[\frac{1+\slash{v}}{2} \bar \Gamma \Gamma\right]+
\frac{m_c^2}{(\np p)^2}\frac{1}{2}{\rm tr}
\left[\frac{1+\slash{v}}{2}\bar\Gamma 
\frac{\slash{n}_+}{2}\Gamma\right]\right)S(u)\nl &&
-\frac{m_c^2}{(\np p)^2}\frac{1}{2}{\rm tr}
\left[\frac{1+\slash{v}}{2}
\bar\Gamma\frac{\slash{n}_-}{2}\Gamma\right]t_1(u).
\eea
To proceed further, we use that the Dirac structures for the
semi-leptonic decay in the Standard Model are
\be
\bar\Gamma=\gamma^\mu(1-\gamma_5), 
\qquad \Gamma=\gamma^\nu(1-\gamma_5).
\ee
The correction linear in $m_c$
vanishes for this V-A structure.   
Performing the traces, and including now all terms,
we find that the components of the hadronic tensor are
given by
\bea \label{eq:wis}
W_1&=&\frac{2}{\np p} 
\bigg[\left(1+\frac{u}{\np p}\right)S(u) +
\frac{s(u)}{2 m_b}
\nonumber\nl
&& -\, \frac{
u S(u)-t(u)}{m_b}
-\frac{u_s(u)+u_a(u)}{\np p} 
-\frac{m_c^2 }{(\np p)^2}t_1(u)\bigg],
\nonumber\\
W_2 &=&\frac{1}{2} W_3 =
-\frac{2 u}{\np p}\,S(u),
\nonumber\\
W_4&=&-\frac{4}{(\np p)^2}
\,t(u),
\nonumber\\
W_5&=& \frac{8}{(\np p)^2}
\bigg[
\frac{u S(u)-t(u)}{m_b}+\frac{
t(u)+u_a(u)}{\np p}\bigg].
\eea
The appearance of the effective shape function $t_1(u)$ within
the  tree approximation is 
specific to decay into charm quarks, as is seen by setting
$m_c\to 0$. This shows that the effects of the final-state 
charm-quark mass are not merely kinematical, 
but introduce new non-perturbative structure into the
factorization formula. This new structure is relevant only
in the shape-function region.  In Section \ref{sec:locexpans}
we will extrapolate our results to the OPE region 
by performing a moment expansion, and find that the  moments 
of $t_1(u)$ are determined entirely by those of the 
leading order shape function 
$S(u)$, so that in this region the effects of the quark
mass are purely kinematical.


\section{Differential Decay Spectra}\label{sec:spectra}

From the scalar components $W_i$ of the hadronic tensor one
can express the triply differential decay rate as \cite{DeFazio:1999sv} 
\bea \label{eq:tripdif}
&&\frac{1}{12\Gamma_0}
\frac {d^3\Gamma}{d(\nm \hat p) d(\np \hat p)d\bar x} =\\
&&(\np \hat p -\nm \hat p)\bigg\{ 
(1+\bar x-\nm\hat p -\np\hat p)
(\np \hat p +\nm \hat p -\bar x - \np \hat p \, \nm \hat p)\frac{m_b^2}{2}
W_1 \nonumber \\
&&+ (1-\np \hat p  + \np \hat p \, \nm \hat p)\frac{m_b}{2}W_2
+[\bar x(\np \hat p  -\bar x)- \np \hat p \, \nm \hat p]
\frac{m_b}{4}(W_3 + 2 m_b W_4 + m_b^2 W_5)\bigg\},
\nonumber
\eea
where $\Gamma_0=G_F^2 m_b^5 |V_{cb}|^2/192\pi^3$ and
$\bar x=1 -2 E_{\ell}/m_b$.  Here and below hatted 
quantities are normalized to $m_b$, i.e.
$\hat p=p/m_b,\, \hat m_c =m_c/m_b$.
We have used that only $W_1$ contains a leading order term in
$\lambda$ to expand the differential decay rate to 
${\cal O}(\lambda^2)$. The integration range is given by
\be
\frac{\hat{m}_c^2}{ \np \hat p}\leq 
\nm \hat p \leq \bar x \leq \np \hat p \leq 1.
\ee
Any decay distribution can be obtained
from (\ref{eq:tripdif}) along with the result for the hadronic
tensor in (\ref{eq:wis}).  Of special interest for semi-leptonic 
$b\to u$ transitions is the singly differential spectrum in
the variable $P_+ = \nm p + (M_B-m_b)$,
 because of the possibilities
to extract $|V_{ub}|$\cite{NeubertP+}.  
An examination of (\ref{eq:wis}) shows
that $u=\nm p - m_c^2/\np p$ is the 
natural partonic analog of $P_+$ for $b\to c$ decay , 
since this is what appears in the argument 
of the shape functions. (We leave the discussion of 
an appropriate hadronic variable to 
Section~\ref{sec:leading}.)
Changing variables
and performing the integrals over $\np \hat p$ and $\bar x$, we find
\bea \label{eq:uspec}
\frac{1}{\Gamma_0}\frac{d \Gamma}{d u}&= &
\left(  1-\frac{14}{3}\frac{u}{m_b}-8 \, \frac{m_c^2}{m_b^2}\right)S(u)
+\frac{s(u)}{2m_b}- 4 \, \frac{m_c^2}{m_b^2} \, t_1(u)\nl
&&+\frac{1}{3 m_b}[t(u)+u_a(u)-5u_s(u)],
\eea
where the allowed phase space is
$0\leq  u\leq m_b- m_c^2/m_b$. This result
agrees with \cite{Bosch:2004cb} in the
limit $m_c\to 0$. 

A comment is in order concerning the limits of integration used 
in calculating (\ref{eq:uspec}).
After changing variables, the integration region is given by
\be
\hat u+\frac{\hat{m}_c^2}{\np\hat p} \leq \bar x \leq \np \hat p, 
\qquad
\frac{\hat u}{2}+\frac12 \sqrt{\hat{u}^2 + 
4 \hat{m}_c^2}\leq \np \hat p \leq 1.
\ee
The collinear expansion is valid in the region $\np \hat p \sim O(1)$,
and breaks down for the lower limit above, which corresponds
to $\np \hat p\sim \hat m_c \sim O(\lambda)$.  However, the 
doubly differential rate behaves like $\lambda^2$ in this
region of phase space.
Hence the contribution to the $u$ spectrum from this 
region is in total of order $\lambda^3$ and does not affect
our analysis.

We can also derive the lepton energy spectrum in the endpoint region,
where an experimental cut limits the lepton-energy 
variable $\bar x$ to values satisfying $\bar x \sim {\cal O}(\lambda^2)$.
Such a cut restricts the other two kinematic variables 
to the shape-function region, since
$\hat m_c^2/\np \hat p \leq \nm \hat p \leq \bar x$
and $\hat m_c^2/(\bar x -\hat u)\leq \np \hat p \leq 1$
ensure that $\nm \hat p \sim {\cal O}(\lambda^2)$ 
and $\np \hat p \sim {\cal O}(1)$.
Integrating over $\np \hat p$, we find that the leading-order result for the
doubly differential spectrum is
\be \label{eq:endspec}
\frac{1}{\Gamma_0}\frac{d^2\Gamma}{d\bar x d u} \approx
2 S(u)\left[1-\frac{3 \hat m_c^4}{(\bar x -\hat u)^2}+
\frac{2 \hat m_c^6}{(\bar x -\hat u)^3}\right] + {\cal O}(\lambda^2).
\ee
We have also calculated the ${\cal O}(\lambda^2)$
power-suppressed terms but do not quote them here.

\section{Comparison with the local expansion}\label{sec:locexpans}

A more familiar treatment of inclusive $b\to c$ decay
uses the operator product expansion (OPE) to calculate the
total rate \cite{Manohar:1993qn}.
Within this approach, the singly differential
spectrum in the variable $u$ at leading order in $\alpha_s$ and
including power-suppressed
corrections up to dimension 5 is given by
\bea\label{eq:OPEu}
\frac{1}{\Gamma_0}\frac{d\Gamma}{du}&=&
\left(1- 8 \, \frac{m_c^2}{m_b^2}\right)\delta(u) \nonumber \\
&&-\left(\frac{17}{18}\frac{\lambda_1}{m_b}+
\frac{3}{2}\frac{\lambda_2}{m_b}\right)\delta^\prime(u)+
\lambda_1\left(\frac{8 m_c^2}{3m_b^2}-\frac{1}{6}\right)
\delta^{\prime\prime}(u)+ {\cal O}\left(\frac{1}{m_b^3}\right).
\eea  
We have not written terms proportional to 
$\lambda_{1,2}\,\delta(u)$, since these are 
$\Lambda^2/m_b^2$ 
corrections which  cannot be recovered
from our calculation in the shape-function region, 
which was carried out only to  
${\cal O}(\Lambda/m_b, \,m_c^2/m_b^2)$.
The OPE result is valid in the kinematical region where all
momentum components are of order  $m_b$.  
We can compare the results calculated in the OPE region
with those derived in the shape-function region by 
considering $u\sim m_b$,
in which case moments of the 
effective shape functions can be expressed
in terms of local HQET matrix elements \cite{NeubertSF}
(see also \cite{Lee:2004ja, Bosch:2004cb, Bauer:2001mh}).
This expansion is given by
\bea\label{eq:moments}
S(u)&=&\delta(u)-\frac{\lambda_1}{6}\delta^{\prime\prime}(u),\qquad
s(u) =-\left(\lambda_1+3\lambda_2\right)\delta^\prime(u),
  \nonumber \\
u_s(u)&=& -\frac{2\lambda_1}{3}\delta^\prime(u), \qquad\qquad\!\!\!
u_a(u)= \lambda_2 \,\delta^\prime(u), \nonumber \\
t(u)&=& -\lambda_2 \,\delta^\prime(u),\qquad\qquad \,
t_1(u)=-\frac{\lambda_1}{3}\delta^{\prime\prime}(u).
\eea
The higher moments correspond to operators of higher than
dimension 5 in the OPE and are neglected.  

We will derive
the moment expansion for $t_1(u)$ as an explicit example, since
it is unique to decay into charm quarks. Examining
the definition (\ref{eq:effshape}), we see that it is sufficient
to derive the moments of the function $T_1(\omega_1,\omega_2)$, since
those of $S(\omega)$ are known. At dimension 5, 
we only need to consider up to the first
moment with respect to $\omega_1$ or $\omega_2$. 
Moments of the form  
$\int d\omega_1 d \omega_2 \, T_1(\omega_1,\omega_2) \, \omega_1^n$
give a vanishing contribution to $t_1(u)$, as is seen by inserting
$T_1(\omega_1,\omega_2)\sim \delta^{(n)}(\omega_1)\delta(\omega_2)$ 
into  (\ref{eq:effshape}). This means that we only have
to calculate the first moment with respect to $\omega_2$. To do this,
we start with its definition in terms of a tri-local
matrix element (see (\ref{YDs}) and (\ref{35}))
\bea\label{eq:defT}
 \frac{n_{+}^\mu}{2}\langle\bar B| 
(\bar h_v Y)(x)[Y^\dagger iD_{\mu}^s Y](z)
(Y^\dagger h_v)(0)|\bar B\rangle =\int d\omega_1 
d\omega_2 e^{-i\omega_1 x_+-i\omega_2 z_+}
 T_1(\omega_1,\omega_2),
\eea
where the derivative does not act outside the square bracket.   
The first moment with respect to $\omega_2$ is obtained by acting on
both sides of (\ref{eq:defT}) 
with $-i\nm \partial_{z}$ 
and then setting $x$ and $z$ to zero to obtain a local HQET
matrix element, which is calculated using
\be
\langle\bar B|\bar h_v \Gamma_{\mu\nu}iD^\mu i D^\nu h_v|\bar B\rangle=
\frac{1}{2}{\rm tr}\left(\Gamma_{\mu\nu}\frac{1+\slash{v}}{2}\left[
(g^{\mu\nu}-v^\mu v^\nu)\frac{\lambda_1}{3} + i \sigma^{\mu\nu}
\frac{\lambda_2}{2}\right]\frac{1+\slash{v}}{2}\right).
\ee
We then find 
$\int d\omega_1 d\omega_2 T_1(\omega_1,\omega_2)\omega_2
 =0$, as 
 can be seen by using the identity
\be
-i\nm\partial_z [Y^\dagger iD_{\mu} Y](z)=  n_{-}^\nu[Y^\dagger
i g F_{\mu\nu} Y](z),
\ee
and noting that the local HQET matrix element vanishes.  
We conclude that $T_1(\omega_1,\omega_2)$ is irrelevant for the moment 
expansion of $t_1(u)$.  Inserting the moment expansion
for $S(\omega)$ into (\ref{eq:effshape}), we find
$t_1(u)=-\frac{\lambda_1}{3} \delta^{\prime\prime}(u)+\dots$,
as quoted above.

Replacing the shape functions in (\ref{eq:uspec}) by the moments
in  (\ref{eq:moments}), we find that the local expansion of the
shape-function region result  reproduces  the
OPE result (\ref{eq:OPEu}). We performed the same comparison for 
the lepton energy spectrum (\ref{eq:endspec}), 
including power-suppressed terms, and confirmed that the moment 
expansion reproduces the OPE result \cite{Manohar:1993qn},
expanded appropriately for $\hat m_c^2 \sim \bar x\sim \lambda^2$.

\section{Leading-Order Shape-Function Effects}\label{sec:leading}

Neglecting sub-leading shape functions and perturbative corrections, the
$d\Gamma/du$ spectrum is directly proportional to the 
leading-order shape function $S(u)$.
This raises the prospect of extracting information on 
the shape function from data on inclusive charm decays, 
a question which we will now address.  Before performing
any phenomenological studies we will slightly modify our
treatment of $m_c^2$ corrections.  In (\ref{eq:uspec}), we kept
only terms up to $m_c^2$, as is consistent with the SCET
power counting.  However, higher-order 
kinematic corrections involving the charm-quark mass arise from
phase-space integrals and are easily included.
Keeping these additional kinematic
corrections, 
but neglecting all other sub-leading effects,
the spectrum becomes
\be\label{eq:leadingu}
 \frac{1}{\Gamma_0}\frac{d \Gamma}{d u}= f( \rho)S(u),
\ee
where $f(\rho)$ is the usual phase-space function
\be \label{eq:ffull}
f(\rho)=
\left(1-8  \rho
+8 \rho^3- \rho^4- 12  \rho^2 \ln \rho \right),
\ee
and we have defined $\rho=m_c^2/m_b^2$.  The numerical effect of 
the phase-space function is sizable. In particular, using
$f(\rho) \sim
\left(1-8 \rho \right)$
would differ at $\rho \sim 0.1$ significantly from the value obtained from 
(\ref{eq:ffull}). However, once the logarithmic contribution is  included, 
the numerical result is almost exact.

In an experimental analysis one has to rewrite (\ref{eq:leadingu}) 
in terms of hadronic variables. To this end, we propose 
to study the spectrum in the variable 
\be
U=\nm P -\frac{M_D^2}{\np P},
\ee
where $n_{\pm}P=n_{\pm}p+(M_B-m_b) = n_{\pm}p+\bar\Lambda$ 
are the hadronic light-cone variables. 
Since $U$ depends on $\np P$ one
must make this replacement before integrating over $\np P$, in which case
the integration region is
\bea \label{eq:phasespace}
\frac{U}{2}+\frac12 \sqrt{U^2+4 M_D^2}&\leq& \np P \leq M_B, \qquad
0 \leq U\leq
 M_B-\frac{M_D^2}{M_B}
\eea
One has to be careful when writing the spectrum in terms of $U$ 
because
\bea
u&=& U-\bar \Lambda + 
\frac{M_D^2}{\np P} -\frac{m_c^2}{\np P-\bar \Lambda} \nonumber \\
&\approx & U-\bar \Lambda + \frac{M_D^2-m_c^2}{\np P}
-\frac{m_c^2}{\np P}\frac{\bar \Lambda}{\np P}+\dots
\eea
The last two terms are of order $\lambda^3$ and $\lambda^4$ 
respectively as long as $\np P\sim M_B$, but become order
$\lambda^2$ at the lower limit of the $\np P$ integration,
when $\np P\sim M_D$. However, one can show that the contribution
of this region to the $U$ spectrum is power-suppressed, 
so we can ignore this subtlety and 
write $u=U-\bar\Lambda$.  The shape function is 
then a function of  $S(U- \bar\Lambda)$, and the 
decay distribution is simply
\be\label{eq:leadingU}
 \frac{1}{\Gamma_0}\frac{d \Gamma}{dU}=  f(\rho) 
S(U-\bar \Lambda).
\ee
At leading order $\rho=m_c^2/m_b^2 \approx M_D^2/M_B^2$.
Eq.~(\ref{eq:leadingU}) opens the possibility to at least cross{}check 
the results for the shape functions obtained from heavy-to-light decays.
However, a few comments are in order.

Unlike in the case of heavy-to-light semi-leptonic decays, almost 80\% 
of the total inclusive rate is already exhausted by the two exclusive
decays $\bar B \to D \ell \bar{\nu}_\ell$ and 
$\bar B \to D^* \ell \bar{\nu}_\ell$. 
This may indicate that the real world is not very close to the limit
$\lambda \to 0$ using the power counting we are suggesting. 
For the exclusive decay $\bar B \to D \ell \bar{\nu}_\ell$ the $U$ spectrum
is concentrated at $U = 0$ 
\be
\frac{d \Gamma (\bar B \to D \ell \bar{\nu}_\ell)}{dU} 
 = \Gamma (\bar B \to D \ell \bar{\nu}_\ell) \, \delta(U)
\ee 
while for the exclusive decay $\bar B \to D^* \ell \bar{\nu}_\ell$ 
the decay rate is a function of 
$$ 
U = \frac{M_{D^*}^2 - M_D^2}{n_+ P} \sim {\cal O} 
(\frac{\lambda^4 M_B^2}{\np P}).
$$ 
Even when $\np P\sim M_D\sim \lambda M_B$ we have
$U\sim \lambda^3 M_B$ and this is beyond the 
sensitivity of the leading-order approach, since
the smearing due to the shape function is of the order $\lambda^2$. 
Thus the contribution of both ground states may be written as  
 \be
\frac{d \Gamma (\bar B \to D \ell \bar{\nu}_\ell)}{dU} + 
\frac{d \Gamma (\bar B \to D^* \ell \bar{\nu}_\ell)}{dU}
 = [ \Gamma (\bar B \to D \ell \bar{\nu}_\ell)  
+ \Gamma (\bar B \to D^* \ell \bar{\nu}_\ell)]
  \, \delta(U).
\ee 
In terms of a moment expansion of the decay rate, we 
see that these two exclusive decay modes contribute only
to the zeroth moment, and conclude that all  
higher moments of the $U$ spectrum can originate only 
from excited or non-resonant states. For this reason,  
it would be interesting to see if and how well the relations
for the moments, such as  
\be\label{eq:moment1}
 \int dU \, U \, \frac{1}{\Gamma_0}\frac{d \Gamma}{dU}=  
  \bar\Lambda  f(\rho) \,  ,
  \qquad 
   \int dU \, U^2 \, \frac{1}{\Gamma_0}\frac{d \Gamma}{dU}=  
  \left( -\frac{1}{3} \lambda_1 + \bar\Lambda^2 \right)  f(\rho)
\ee
are satisfied. In particular, we find that to leading order in the SCET expansion
the ratio of moments should be the same as the corresponding ratio of $P_+$ 
moments in $\bar B \to X_u \ell \bar{\nu}_\ell$ or of photon energy moments in 
$\bar B \to X_s \gamma$.   

\section{Conclusions}\label{sec:conclusions}

In this work we examined inclusive semi-leptonic $b\to c$ decay
using the power counting $m_c\sim \sqrt{\Lambda_{\rm QCD} m_b}$
for the charm-quark mass. 
With this power counting the
decay kinematics can be chosen to access the shape-function
region even in the decay $\bar B \to X_c \ell \bar{\nu}_\ell$.  
We applied effective field theory methods by
modifying SCET to include a massive collinear charm quark.
This defines a consistent power-counting scheme in terms of the
small parameter $\lambda \sim m_c/m_b \sim \sqrt{\Lambda_{\rm QCD}/m_b}$. 
We matched the Lagrangian as well as the weak transition current
at tree level and including sub-leading terms up to  ${\cal O}(\lambda^2)$, 
and used these to derive the hadronic tensor in the shape-function region. 

The results are similar to those for $b\to u$ transitions.
This led us to identify a certain variable 
which is analogous to the variable $P_+$  in $b\to u$ decay.  
As in the heavy-to-light case, at leading power and at tree level
the singly differential spectrum in this variable is  directly proportional 
to the leading-order shape function. We showed how
the moment expansion of this differential spectrum matches
the local OPE result in an appropriate limit. 
 
It remains to be checked whether the power counting we have applied 
for the charm mass yields a consistent picture or whether the 
conventional counting $m_c\sim m_b$ is more appropriate. 
A cross{}check can be performed by measuring the moments of the distribution 
in this variable and by comparing them to those obtained
from the $P_+$ spectrum in $\bar B \to X_u \ell \bar{\nu}_\ell$ and from the 
$E_\gamma$ spectrum in $\bar B \to X_s \gamma$. Our leading-order 
prediction is that these moments should be equal up to the phase-space 
factor which appears in the total partonic rates.

\section*{Acknowledgements}
This work was supported by the DFG Sonderforschungsbereich SFB/TR09 
``Computational Theoretical Particle Physics''.

\end{document}